\documentclass[conference]{IEEEtran} 		
\IEEEoverridecommandlockouts
\pdfoutput=1
\usepackage{ushort}
\usepackage[mathcal]{euscript}
\usepackage{amsmath}
\usepackage{amsthm}
\usepackage{amssymb}
\usepackage{graphicx}
\usepackage{epsfig}
\usepackage{epstopdf}
\usepackage{booktabs}
\usepackage{colortbl}
\usepackage{amsmath}
\usepackage{xcolor}
\usepackage{graphicx}
\usepackage{algpseudocode}
\usepackage{algorithmicx}
\usepackage{algorithm}
\usepackage{cite}
\usepackage{url}
\usepackage{rotating}
\usepackage{array}

\newtheorem{theorem}{Theorem}

\def \apu {\lambda_{1}}
\def \bed {\beta_{2}}
\def \bpu {\beta_{1}}

\def \ued {U_{2}}
\def \upu {U_{1}}

\def \led {\tilde{l}_{2}}
\def \lpu {\tilde{l}_{1}}
\def \ced {c_{2}}
\def \cpu {c_{1}}
\def \ded {d_{2}}
\def \dpu {d_{1}}
\def \Bed {b_{2}}
\def \Bpu {b_{1}}
\def \Aed {a_{2}}
\def \Apu {a_{1}}

\def \t {\text}
\def \te {\text{e}}

\begin{document}

\title{A Delay-Optimal Packet Scheduler for M2M Uplink}

\author{\IEEEauthorblockN{Akshay Kumar, Ahmed Abdelhadi and Charles Clancy}
\IEEEauthorblockA{Hume Center, Virginia Tech\\
Email:\{akshay2, aabdelhadi, tcc\}@vt.edu}
\thanks{This research is based upon work supported by the National Science Foundation under Grant No. 1134843. 
}
}

\maketitle
\thispagestyle{plain}
\pagestyle{plain}
\begin{abstract}
In this paper, we present a delay-optimal packet scheduler for processing the M2M uplink traffic at the M2M application server (AS). Due to the delay-heterogeneity in uplink traffic, we classify it broadly into delay-tolerant and delay-sensitive traffic. We then map the diverse delay requirements of each class to sigmoidal functions of packet delay and formulate a utility-maximization problem that results in a proportionally fair delay-optimal scheduler. We note that solving this optimization problem is equivalent to solving for the optimal fraction of time each class is served with (preemptive) priority such that it maximizes the system utility. Using Monte-Carlo simulations for the queuing process at AS, we verify the correctness of the analytical result for optimal scheduler and show that it outperforms other state-of-the-art packet schedulers such as weighted round robin, max-weight scheduler, fair scheduler and priority scheduling. We also note that at higher traffic arrival rate, the proposed scheduler results in a near-minimal delay variance for the delay-sensitive traffic which is highly desirable. This comes at the expense of somewhat higher delay variance for delay-tolerant traffic which is usually acceptable due to its delay-tolerant nature.
\end{abstract}


\section{Introduction}
Over the past few years, the global Machine-to-Machine (M2M) communications has witnessed a significant growth due to its multitude of industrial use-cases such as industrial automation, Smart Grid, Digital billboards etc \cite{M2Mgrowth}. The M2M uplink is of more practical interest than the downlink as the typical use-case consists of device monitoring and control functions. M2M devices generate sporadic data with small payload size, typically few hundred bytes and therefore, the focus is on availability and reliability as opposed to the high data rate requirement of human-oriented-services. The traffic arrival rate, payload size and delay requirements of M2M uplink traffic differ vastly for different M2M use-cases. For instance, consider the smart-grid traffic \cite{smartGrid} originating from the smart meters in residential homes. As per the UCA OpenSG specification (described in \cite{openSG}), the meter reading reports are delay-tolerant with maximum latency requirement greater than $60$~s, average payload of $1.2$~kB and arrival rate of $6$ messages/day/device. On the other extreme is the delay-sensitive real-time pricing messages with latency requirement of less than $3$~s, payload of less than $25$~B and arrival rate of $96$ messages/day/device \cite{openSG}. This extreme heterogeneity in traffic characteristics makes it is imperative to design delay-aware packet schedulers for serving the M2M uplink traffic \cite{Maia14}.

Most of the existing M2M packet schedulers are designed for specific wireless technology such as LTE and are heuristic schedulers (see \cite{Gotsis12} and references therein). Most of the LTE packet schedulers use some variants of Access Grant Time Interval scheme for allocating fixed or dynamic access grants over periodic time intervals to M2M devices. Another line of work focuses on optimal scheduling algorithms specifically for real-time embedded systems (see \cite{Buttazzo11} and references therein). The drawback of these schemes is that they assume the arrival and service time of packets are known \emph{apriori}. Therefore, they cannot be used for scheduling M2M traffic which is typically random with different delay requirements for different traffic classes.

 Recently, packet scheduling heuristics that incorporate the heterogeneity in M2M uplink using utility functions have been proposed in \cite{sysCon16arxiv, milcom16arxiv}. However, these schemes are heuristics and do not provide any guarantees on their convergence or delay-performance. Besides this, a number of state-of-the-art packet schedulers exist for queuing networks that can be used for scheduling in M2M uplink. Fair queuing \cite{Demers89, Greenberg90} provides perfect fairness among different traffic classes but almost all its implementations suffer from high operational complexity and also do not account for diverse delay requirements of traffic. To get around these issues, weighted round robin (WRR) and weighted fair scheduling \cite{Parkeh93} have been proposed that sacrifice certain degree of fairness to incorporate the traffic heterogeneity in their design of weights. However the determination of the delay-optimal weights  is not easy and are assigned based on certain criteria chosen by the site administrator. Tassiulas et. al. \cite{Tassiulas} proposed the throughput-optimal max-weight scheduling algorithm, but it results in highly unfair resource allocation when the traffic arrival rates are skewed.

In this paper, we propose a online delay-optimal packet scheduler for M2M uplink. We classify the M2M traffic aggregated at M2M application server (AS) into two broad classes consisting of delay-sensitive and delay-tolerant traffic, with potentially different payload size and packet arrival rates. We then employ sigmoidal functions to map the traffic delay requirements onto utility functions (of packet delay) for each class. The sigmoidal function is versatile enough to represent any arbitrary delay requirements, by appropriately modifying its parameters. Our goal is to determine the optimal scheduling policy that maximizes a proportionally fair system utility metric. We note that, for any given scheduler, the average delay of a class can be expressed as a convex combination of its maximum delay (when served with least preemptive priority) and minimum delay (when served with highest preemptive priority\footnote{Hereafter, we drop the qualifier \lq preemptive\rq~for succinctness.}). Thus the average delay of classes for any work-conserving\footnote{A work-conserving 
scheduler is the one that does not go idle when there are jobs waiting to be served.} scheduler can be realized by appropriately time sharing between different priority policies. Hence, the proposed optimal scheduler is determined by solving for the optimal fraction of time-sharing between the two priority policies that maximizes the system utility. Using Monte-Carlo simulations, we verify the correctness of the proposed scheduler and show that it performs better than various state-of-the-art schedulers such as WRR, max-weight scheduler, fair scheduler and priority scheduling. We also note that at higher traffic arrival rate, the proposed scheduler results in a much smaller delay variance for the delay-sensitive traffic as compared to the other schedulers. Lastly the proposed scheduler is agnostic to communication standard used for M2M uplink and easily adapt to time-varying characteristics of M2M traffic.

The rest of the paper is organized as follows. Section~\ref{sysModel} introduces the system model for M2M uplink. Then in Section~\ref{probForm}, we map the traffic delay requirements onto the utility functions, formulate the utility maximization problem and then present the optimal scheduler. Section~\ref{Results} presents simulation results. Finally, Section~\ref{concl} draws some conclusions.

\section{System Model}
\label{sysModel}
\begin{figure}%
\centering
\includegraphics[width = \columnwidth]{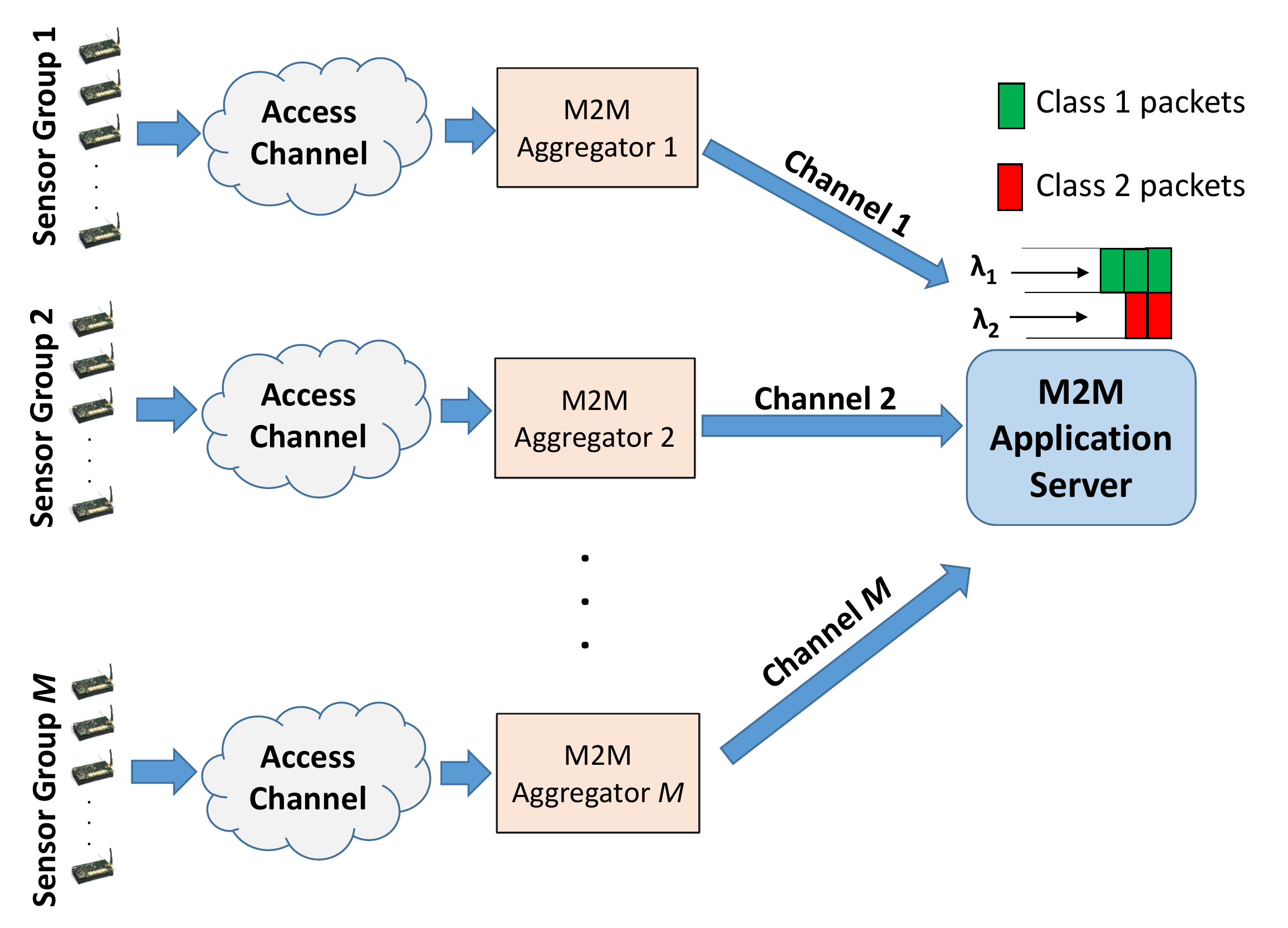}%
\vspace{0pt}
\caption{M2M uplink model illustrating the queuing process at M2M Application Server.}%
\label{systemModel}%
\end{figure}

Fig.~\ref{systemModel} shows the system model for a generic M2M uplink. The sensory data from each local group of sensors is first aggregated at a M2M Aggregator (MA) and then the data from multiple MA's is aggregated and queued at a M2M Application Server (AS). The data from each MA is transmitted to the AS using parallel orthogonal channels\footnote{Using Orthogonal Frequency Division Multiple Access, the data from each MA-AS link is assigned a set of orthogonal sub-carriers. The number of sub-carriers assigned to each MA-AS link depends on its data-rate requirements.}. Since the M2M traffic from multiple sensors is aggregated at each MA, assigning dedicated resources to each MA-AS link does not result in any significant resource wastage.

We assume that the data from the sensors can be broadly classified into two classes at A: class $1$ for delay-sensitive and class $2$ for delay-tolerant traffic. We assume that the arrival process for class $i$ is Poisson with rate $\lambda_i$ \cite{Dhillon14}. Consider a general packet size distribution for class $i$ with average packet size of $s_i$. Let $r$ denotes the AS service rate and $X_i$ be the resultant general service time distribution for class $i$ with service rate $\mu_i=r/s_i$. 
 The total time spent by a packet in the system, $T$, can be written as sum of following components,
\begin{equation}
T = T_{\text{trans}}+T_{\text{prop}}+T_{\text{cong}}+T_{\text{queue}}+T_{\text{ser}},
\label{sojTime}
\end{equation}
which denote the following component delays:
\begin{itemize}
	\item $T_{\text{trans}}$: Transmission delay at the sensor.
	\item $T_{\text{prop}}$: Propagation delay from the sensor to MA and from MA to AS.
	\item $T_{\text{cong}}$: Congestion delay due to shared wireless channel in large-scale sensor network.
	\item $T_{\text{queue}}$: Queuing delay at the AS.
	\item $T_{\text{ser}}$: Processing time for a packet at the AS.
\end{itemize}

The small packet size of M2M data relative to transmission rate allows us to ignore $T_{\text{trans}}$. Also, the propagation time $T_{\text{prop}}$ is quite small relative to other delay components and thus can be safely ignored. The congestion delay $T_{\text{cong}}$ for the sensor-MA link can be ignored due to small number of sensors under each MA, each with low traffic rate. The congestion delay for MA-AS links is ignored due to the assumption of dedicated channel for each link. Therefore in this work, we ignore all the terms except the queuing delay, $T_{\text{queue}}$ and the service time $T_{\text{ser}}$ at AS.

Now the queuing delay for each class at AS depends upon the scheduling policy adopted at the AS. The scheduling policy at AS should be chosen as to maximally satisfy the latency constraints of packets of both classes.	

\section{Problem Formulation}
\label{probForm}
In this section, using sigmoidal function \cite{AbdelhadiCNC2014, AbdelhadiPIMRC2013}, we first map the delay requirements for class $i$ onto its utility function as,
\begin{align}
U_i(l_i) = 1-c_i\left(\frac{1}{1+{\text{e}}^{-a_i(l_i-b_i)}}-d_i \right), i=1,2,
\label{utilSig}
\end{align}
where, $c_i=\frac{1+{\text{e}}^{a_ib_i}}{{\text{e}^{a_ib_i}}}$, $d = \frac{1}{1+{\text{e}^{a_ib_i}}}$ and $l_i$ is the latency\footnote{We use the terms \lq delay\rq and \lq latency\rq~interchangeably in this work.} of a class $i$ packet. Note that $U(0)=1$ and $U(\infty) = 0$. The parameter $a_i$ is the utility roll-off factor for class $i$ and the inflection point for the utility occurs at $l_i = b_i$.

\begin{figure}[!t]
\centering
\includegraphics[height = 2.4 in, width = 3.2 in]{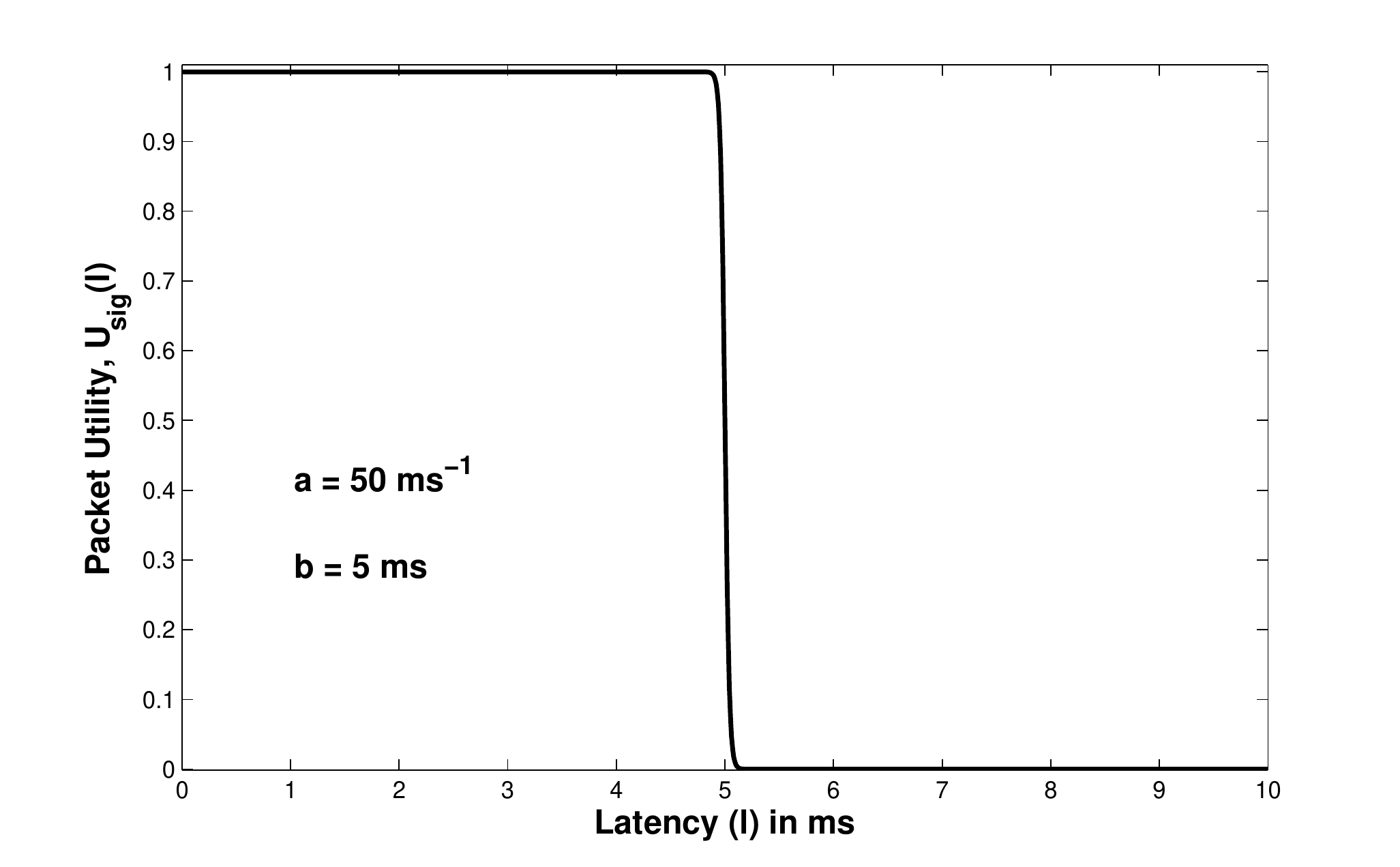}   
\caption{Utility function for class $1$ with $a_1 = 50~\text{ms}^{-1}$ and $b_1 = 5$~ms. }
\label{puUtilFcn} 
\end{figure}

\begin{figure}[!t]
\centering
\includegraphics[height = 2.4 in, width = 3.2 in]{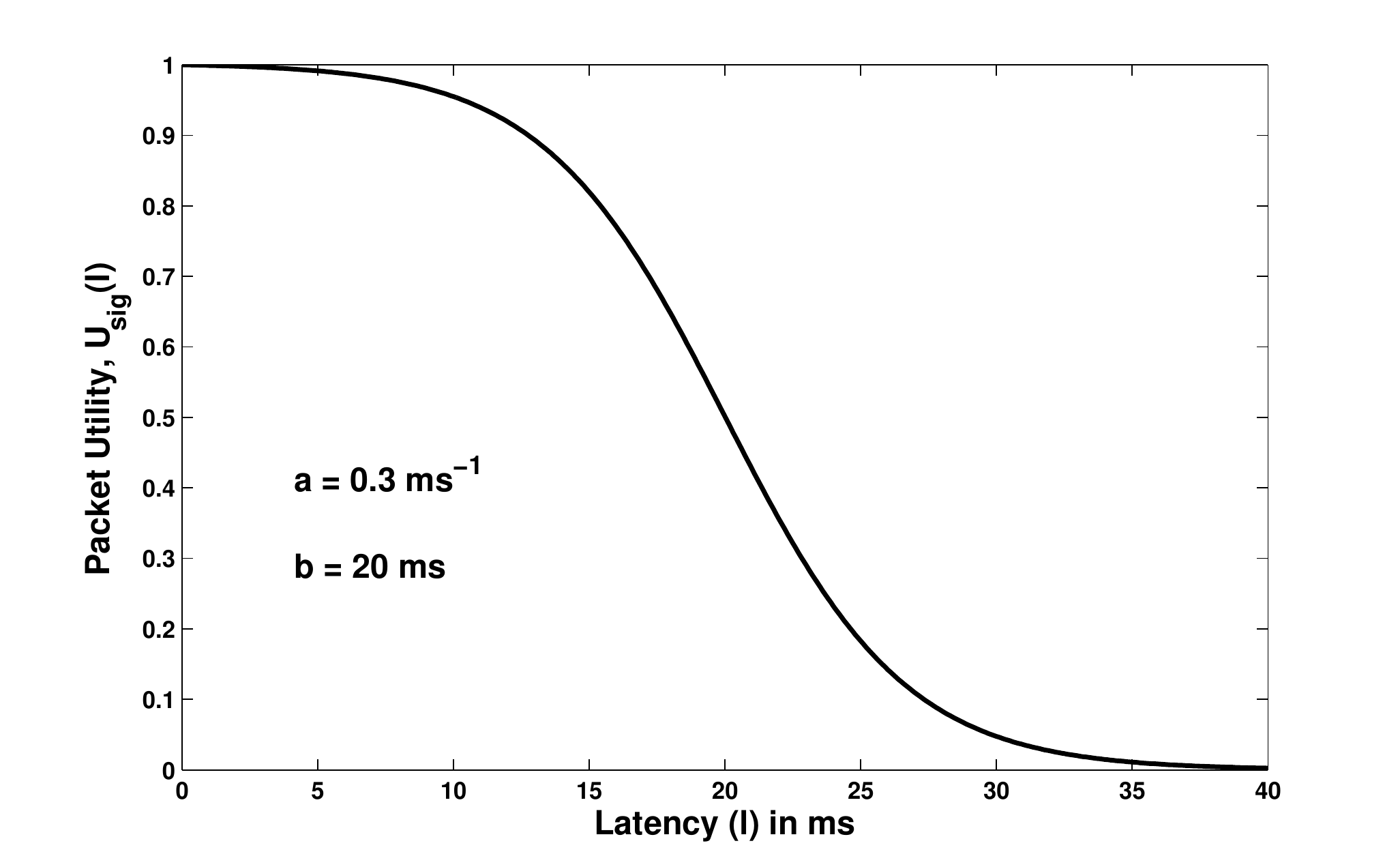}%
\caption{Utility function for class 2 with $a_2 = 0.3~\text{ms}^{-1}$ and $b_2 = 20$~ms.}
\label{edUtilFcn} 
\end{figure}

The sigmoidal function is versatile to represent diverse delay requirements by appropriately changing its parameters. For high $a$ and low $b$, the utility function becomes \lq brick-walled\rq~ (see Fig.~\ref{puUtilFcn}) and is a good fit for delay-sensitive traffic of class $1$. On the other hand, at low value of $a$ and high $b$, the sigmoidal utility function is a good fit for delay-tolerant traffic of class $2$ as shown in Fig.~\ref{edUtilFcn}. 

\subsection{System utility function} For a given scheduling policy $\mathcal{P}$, we first define a proportionally fair system utility function as,
\begin{equation}
V(\mathcal{P}) = {\bold{\upu}}^{\bpu}(\mathcal{P})*{\bold{\ued}}^{\bed}(\mathcal{P}),
\label{sysUtil}
\end{equation}
where $\bold{U_i}(\mathcal{P}) = U_i(\tilde{l}_i)$ is the utility for class $i$ in steady state. $\tilde{l}_i$ is the average latency of class $i$ in steady state and can be expressed as,
\begin{equation}
\label{expAvgUtil}
\tilde{l}_i = \lim_{{T_{\text{s}}} \rightarrow \infty} \frac{\sum_{j=1}^{M_{i}(T_{\text{s}})} l_i^j(\mathcal{P})}{M_i(T_{\text{s}})}, i=1,2,
\end{equation}
where $M_i(T_{\text{s}})$ is the number of packets of class $i$ served in time $T_{\text{s}}$ and $l_i^j$ is the latency of the $j^{\text{th}}$ packet of class $i$. The parameters $\bpu$ and $\bed$ denote the relative importance of utilities of two classes, with a higher value indicating higher importance of that class towards the overall system utility.

\subsection{Optimization Problem}
Assume that in a sufficiently large time interval\footnote{We assume the time interval under consideration is large enough to observe steady state queuing behavior.}, the AS serves the class $1$ and class $2$ packets with (preemptive) priority for $\alpha_1$ and $\alpha_2$ fraction of the time respectively. Then the average latency of two classes can be expressed as,
\begin{equation}
\label{latPUED}
\begin{aligned}
\lpu &= \alpha_1 \tilde{l}_{11} + \alpha_2 \tilde{l}_{12},  \\
\led &= \alpha_2 \tilde{l}_{21} + \alpha_1 \tilde{l}_{22}, 
\end{aligned}	
\end{equation}
where $\tilde{l}_{ij}$ denotes the average latency of class $i$ when class $j$ is served with higher priority. We then use the existing results for average latency of a class in M/G/1 priority queuing systems by Bertsekas et. al. \cite{gallager} to get,
\begin{equation}
\label{latEqSet}
 l_{ij} = 
  \begin{cases} 
   \frac{R_{ii}+(1-\rho_i)(1/\mu_i)}{1-\rho_i} & \text{if } i = j, \\   
   \frac{R_{ij}+(1-\rho_1-\rho_2)(1/\mu_i)}{(1-\rho_j)(1-\rho_1-\rho_2)} & \text{if } i \neq j, 
  \end{cases}
\end{equation}
where $\rho_i = \lambda_i/\mu_i$ is the AS utilization ratio for class $i$ and $R_{ij}~(i,j \in \{1, 2\})$ denotes the the average residual time of class $i$ packet when class $j$ has higher priority as given by,
\begin{equation}
\label{resTime} 
 R_{ij} = 
  \begin{cases} 
   \lambda_i \bar{X_i^2}/2       & \text{if } i = j, \\\left(\lambda_1 \bar{X_1^2}+\lambda_2 \bar{X_2^2}\right)/2 & \text{if } i \neq j, 
  \end{cases}
\end{equation}
where $\bar{X_i^2}$ denotes the second moment of distribution $X_i$.
 
Since this scheduling policy is completely characterized by $\bar{\alpha} = [\alpha_1 \alpha_2]$, the system utility in Eq.~\eqref{sysUtil} becomes,
\begin{equation}
V(\bar{\alpha}) = \upu(\bar{\alpha})^{\bpu}*\ued(\bar{\alpha})^{\bed}.
\label{systUtil1}
\end{equation}

Exploiting the strictly increasing nature of logarithms, we formulate the following utility-maximization problem to determine the optimal $\alpha_1$ and $\alpha_2$,
\begin{equation}
\begin{aligned}
&\operatorname*{max}_{\bar{\alpha}} & & \text{log}\left(V(\bar{\alpha})\right) = {\bpu}\text{log}\left(\upu(\bar{\alpha})\right) + {\bed}\text{log}\left(\ued(\bar{\alpha})\right) \\
&\text{s.t.} & &\alpha_1, \alpha_2 \geq 0, \\
& & &\alpha_1+\alpha_2 = 1.
\end{aligned}
\label{optimProb1}
\end{equation}
\begin{theorem}
\label{theorem1}
The optimization problem in Eq.\eqref{optimProb1} is convex.
\end{theorem}

\begin{IEEEproof}
We first prove that $f^i(\bar{\alpha}) = \text{log}\left(U_i(\bar{\alpha})\right)$ is concave $\forall~i$. The Hessian matrix of $f^i$ at $\bar{\alpha}$ is defined as,
\[ H^i(\bar{\alpha}) = \left( \begin{array}{cc}
f^i_{11}(\bar{\alpha}) & f^i_{12}(\bar{\alpha}) \\
f^i_{21}(\bar{\alpha}) & f^i_{22}(\bar{\alpha}) 
\end{array} \right), i=1,2,\]
where $f^i_{jk} = \frac{\partial^2 f^i}{\partial \alpha_j \partial \alpha_k}$. 

On solving, we get the following result,
\begin{equation}
f^i_{jk}(\bar{\alpha})  = \frac{-\beta_i \theta_i a_i^2 \tilde{l}_{i,j} \tilde{l}_{i,k}}{{(1+\theta_i)}^2},
\end{equation}
where $\theta_i = \te^{-a_i(\tilde{l}_i-b_i)}$.

Therefore, $H^i(\bar{\alpha})$ is given by,  $H^i(\bar{\alpha}) =$
\begin{equation}
\frac{-\beta_i \theta_i a_i^2}{{(1+\theta_i)}^2} \!\left( \begin{array}{cc}
\tilde{l}_{i1}^2 & \tilde{l}_{i,1} \tilde{l}_{i,2} \\
\tilde{l}_{i,1} \tilde{l}_{i,2} & \tilde{l}_{i2}^2
\end{array} \right),
\label{hessian}
\end{equation}

Now to prove that $f^i$ is a concave function, it is sufficient to prove that $H^i$ is a Negative Semi-Definite (NSD) matrix for all $\bar{\alpha}$ that satisfies the constraints in \eqref{optimProb1}. Let $\Delta_k$ denote a $k^{\t{th}}$ order principal minor of $H^i$. Then $H^i$ is NSD if and only if $(-1)^k \Delta_k \geq 0$ for all principal minors of order $k = 1, 2$.

From Eq.\eqref{hessian}, we get the principal minors as,
\begin{equation}
\begin{aligned}
\Delta_1 &= \frac{-\beta_i \theta_i a_i^2 \tilde{l}_{i,1}^2}{{(1+\theta_i)}^2}, \frac{-\beta_i \theta_i a_i^2 \tilde{l}_{i,2}^2}{{(1+\theta_i)}^2},\\
\Delta_2& = 0.
\end{aligned}
 \end{equation}
Clearly all $\Delta_1 < 0$ and $\Delta_2 = 0$. Therefore $H^i(\alpha_{\mathcal{A}_r})$ is NSD for all $\bar{\alpha}$. Hence, $f^i(\bar{\alpha})$ is concave for all $\bar{\alpha}$ and $i=1,2$. This implies that the aggregated utility natural logarithm $\text{log}\left(V(\bar{\alpha})\right) = {\bpu}\text{log}\left(\upu(\bar{\alpha})\right) + {\bed}\text{log}\left(\ued(\bar{\alpha})\right)$ is also concave. Lastly, the convexity of the optimization problem in Eq.~\eqref{optimProb1} follows from the concavity of the objective function and the affine nature of the constraints. 
\end{IEEEproof}

\section{Optimal Scheduler}
Having established the convexity of the optimization problem, we now proceed to solve it to determine the optimal solution. To make the analysis simple, we reduce the number of variables by setting $\alpha_1 = \alpha$ and $\alpha_2 = 1-\alpha$ in Eq.~\eqref{optimProb1}. Then the only constraint is that $0 \leq \alpha \leq 1$. We solve the unconstrained optimization problem and then factor in the constraint on $\alpha$ by providing a boundary condition. At the optimal solution $\alpha^{*}$, we have,
\begin{align} 
&\frac{\mathrm{d}}{\mathrm{d}\alpha} \text{log}(V(\alpha)) \Big|_{\alpha = \alpha^{*}} = 0,  \nonumber \\ 
\implies &\frac{\bpu}{\upu(\alpha)} \frac{\mathrm{d}}{\mathrm{d}\alpha} \upu(\alpha) \Big|_{\alpha = \alpha^{*}} + \frac{\bed}{\ued(\alpha)} \frac{\mathrm{d}}{\mathrm{d}\alpha} \ued(\alpha) \Big|_{\alpha = \alpha^{*}} = 0. \label{derivativeSysUtil}
\end{align}

Using Eq.~\eqref{utilSig}, we have,
\begin{equation}
\label{derivPUEDutil}
\begin{aligned}
\frac{\mathrm{d}}{\mathrm{d}\alpha} \upu(\alpha) &= \frac{-\cpu \Apu \te^{-\Apu(\tilde{l}_1-\Bpu)} \left(\tilde{l}_{11} - \tilde{l}_{12}\right)}{{\left(1+ \te^{-\Apu(\lpu-\Bpu)}\right)}^2}, \\
\frac{\mathrm{d}}{\mathrm{d}\alpha} \ued(\alpha) &= \frac{-\ced \Aed \te^{-\Aed(\led-\Bed)} \left(\tilde{l}_{21} - \tilde{l}_{22} \right)}{{\left(1+ \te^{-\Aed(\led-\Bed)}\right)}^2}.
\end{aligned}
\end{equation}

Now using Eq.~\eqref{utilSig} and Eq.~\eqref{derivPUEDutil} in Eq.~\eqref{derivativeSysUtil}, we get,
\begin{align}
\frac{\bpu \cpu \Apu \te^{-\Apu(\lpu-\Bpu)}  (\tilde{l}_{12} - \tilde{l}_{11})} {\left(\left(1+\cpu \dpu \right)\left(1+ \te^{-\Apu(\lpu-\Bpu)}\right)- \cpu \right)\left(1+ \te^{-\Apu(\lpu-\Bpu)}\right)} = \nonumber \\ \frac{\bed \ced \Aed \te^{-\Aed(\led-\Bed)} (\tilde{l}_{21} - \tilde{l}_{22})} {\left(\left(1+\ced \ded \right)\left(1+ \te^{-\Aed(\led-\Bed)}\right)- \ced \right)\left(1+ \te^{-\Aed(\led-\Bed)}\right)}.
\label{fullEqnDeriv}
\end{align}

Noting that $\left(1+\cpu \dpu \right) = \cpu$ and $\left(1+\ced \ded \right) = \ced$, and rearranging the terms we get,

\begin{align}
 \frac{\bpu \Apu  (\tilde{l}_{12} - \tilde{l}_{11})} {\bed \Aed (\tilde{l}_{21} - \tilde{l}_{22})}  = \frac{1+ \te^{-\Apu(\lpu-\Bpu)}} {1+ \te^{-\Aed(\led-\Bed)}}.
\end{align}

\begin{figure}[!h]%
\centering
\includegraphics[height = 2.6 in, width = 3.5 in]{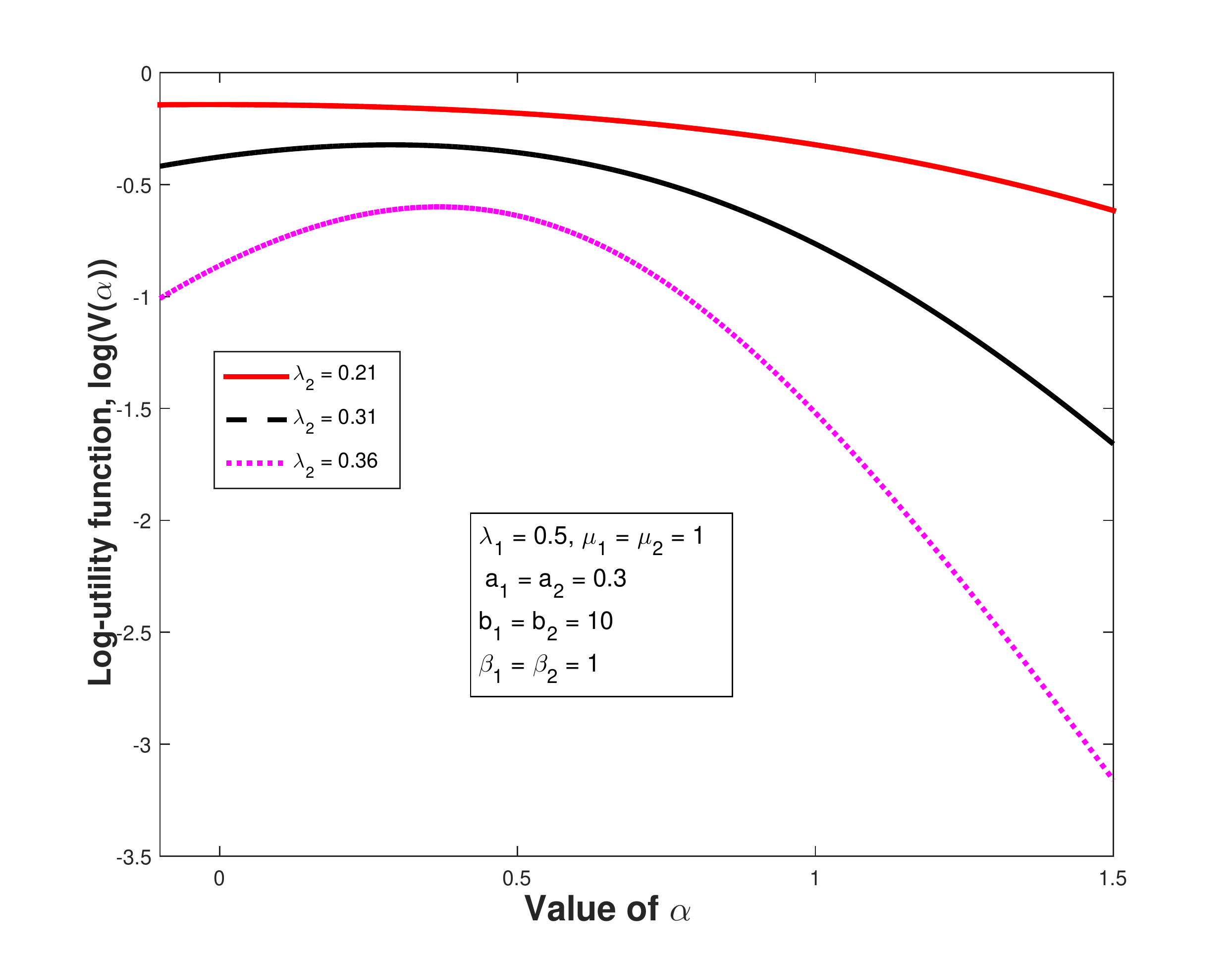}%
\caption{Concavity of the objective function $\text{log}(V(\alpha))$ for $\alpha \in [0,1]$.}%
\label{concaveObj}%
\end{figure}

We note that the LHS is a constant and denote it by $\theta$. Using algebraic manipulations, we get,
\begin{align}
\theta \left(1+ \te^{\Aed(\Bed-\tilde{l}_{22})} \te^{-\Aed \alpha^{*} (\tilde{l}_{21}-\tilde{l}_{22})}\right) &= \nonumber \\ 1+ \te^{\Apu(\Bpu-\tilde{l}_{12})} \te^{-\Apu \alpha^{*} (\tilde{l}_{11}-\tilde{l}_{12})}, \nonumber \\
\implies \phi_{21} \te^{\alpha^{*} \phi_{22}} - \phi_{11} \te^{\alpha^{*} \phi_{12}} + \phi &= 0,
\label{finalSoln}
\end{align}
where we have,
\begin{equation}
\label{constants}
\begin{aligned}
& \phi_{21} = \theta \te^{\Aed(\Bed-\tilde{l}_{22})},~~\phi_{22} =  \Aed (\tilde{l}_{22}-\tilde{l}_{21}),  \\
&\phi_{11} = \te^{\Apu(\Bpu-\tilde{l}_{12})},~~\phi_{12} =  \Apu (\tilde{l}_{12}-\tilde{l}_{11}),  \\
& \phi =  \theta-1.  
\end{aligned}
\end{equation}
Now imposing the constraint $0 \leq \alpha^{*} \leq 1$, we have,
\begin{equation}
 \alpha^{*} = 
  \begin{cases} 
   0      & \text{if}~~ \alpha^{*} < 0 \\
   \alpha^{*} & \text{if }~ 0 \leq \alpha^{*} \leq 1 \\
   1       & \text{if}~~ \alpha^{*} > 1.
  \end{cases}
\end{equation}
This is due to the concave nature of the objective $\text{log}(V(\alpha))$ as shown in Fig.~\ref{concaveObj}. If $\alpha^{*} \notin [0, 1]$, then the maxima lies at the extreme points of the feasible region i.e., we set $\alpha^{*} = 0$ or $1$.

Again due to concavity of $\text{log}(V(\alpha))$, if $\alpha^{*} \in [0, 1]$, then $\frac{\mathrm{d}}{\mathrm{d}\alpha} \text{log}(V(\alpha))$ has opposite signs at $\alpha= \{0,1\}$. Mathematically, we write this as $Z^{'}(0)*Z^{'}(1) < 0$, where $Z(\alpha) = \text{log}(V(\alpha))$ and the superscript ${'}$ indicates the first derivative. If $Z^{'}(0) < 0$ and $Z^{'}(1) < 0$, then set $\alpha^{*} = 0$. If $Z^{'}(0) > 0$ and $Z^{'}(1) > 0$, then set $\alpha^{*} = 1$. After some algebraic simplifications, $Z^{'}(\alpha)$ can be written as, 
\begin{align}
Z^{'}(\alpha) &= \frac{\bpu \Apu (\tilde{l}_{12}-\tilde{l}_{11})} {1+ \te^{-\Apu(\lpu-\Bpu)}} -  \frac{\bed \Aed (\tilde{l}_{21}-\tilde{l}_{22})} {1+ \te^{-\Aed(\led-\Bed)}}.
\end{align}
Therefore, we have,
\begin{equation}
 \label{zprime}
\begin{aligned}
Z^{'}(1) &= \frac{\bpu \Apu (\tilde{l}_{12}-\tilde{l}_{11})} {1+ \te^{-\Apu(\tilde{l}_{11}-\Bpu)}} -  \frac{\bed \Aed (\tilde{l}_{21}-\tilde{l}_{22})} {1+ \te^{-\Aed(\tilde{l}_{21}-\Bed)}},\\
Z^{'}(0) &= \frac{\bpu \Apu (\tilde{l}_{12}-\tilde{l}_{11})} {1+ \te^{-\Apu(\tilde{l}_{12}-\Bpu)}} -  \frac{\bed \Aed (\tilde{l}_{21}-\tilde{l}_{22})} {1+ \te^{-\Aed(\tilde{l}_{22}-\Bed)}}.
\end{aligned}
\end{equation}

The algorithm for the proposed optimal packet scheduler in Algorithm~\ref{propScheduler}.

\begin{algorithm}[h]
\caption{Proposed Delay-Optimal Packet Scheduler}
\label{propScheduler}
		\begin{algorithmic} 
		\label{algoProp}
		\State Compute $\tilde{l}_{11}$, $\tilde{l}_{12}$, $\tilde{l}_{21}$ and $\tilde{l}_{22}$ using Eq.~\eqref{latEqSet}.
		\State Compute  $\phi$, $\phi_{11}$, $\phi_{12}$, $\phi_{21}$ and $\phi_{22}$ using Eq.~\eqref{constants}.
		\State Compute $Z^{'}(1)$ and $Z^{'}(0)$ from Eq.~\eqref{zprime}. 
		\If{$Z^{'}(1)*Z^{'}(0)>0$} 
			\If{$Z^{'}(1)>0$ and $Z^{'}(0)>0$}
				\State Set $\alpha^{*} = 1$. 
			\Else
				\State Set $\alpha^{*} =0$.
			\EndIf
		\Else
			\State Determine $\alpha^{*}$ by numerically solving Eq.~\eqref{finalSoln}. 
		\EndIf	
	\end{algorithmic}	
\end{algorithm}

\section{Simulation Results}
\label{Results}
In this section, we use Monte-Carlo simulations to first evaluate the accuracy of the analytical result for proposed scheduler given in Eq.~\eqref{finalSoln} and then compare the system utility and delay jitter performance of the proposed scheduler against other packet schedulers such as WRR, max-weight scheduler, fair scheduler and priority scheduling. The arrival rate for class $1$ is set to $\apu=0.4~\text{s}^{-1}$. For simplicity, we assume constant packet size for each class and thus $\bar{X_i^2}$ in Eq.~\ref{resTime} is simply $1/\mu_i^2$. We set to $s_1 = s_2 = 100$~B and the service rate is $r=100$ B/s. The utility function parameters for class $1$ are set to $\Apu = 1~\text{s}^{-1}$, $\Bpu = 5$~s and $\bpu = 1$. For class $2$, the utility parameters are $\Aed = 0.3~\text{s}^{-1}$, $\Bed = 10$~s and $\bed = 0.3$. The determination of delay-optimal weights for WRR is non-trivial. However, a sub-optimal yet delay-efficient weight assignment for class $i$ is to set its weight, $w_i$, inversely proportional to its delay requirement (set as $(b_i+\frac{4}{a_i}$ for simplicity) and its packet size $s_i$. After normalizing to integer values, we get the weights as $[w_1, w_2] = [111, 43]$. 

\subsection{System utility performance of different schedulers}
\begin{figure}[!h]%
\centering
\includegraphics[height = 2.5 in, width = \columnwidth]{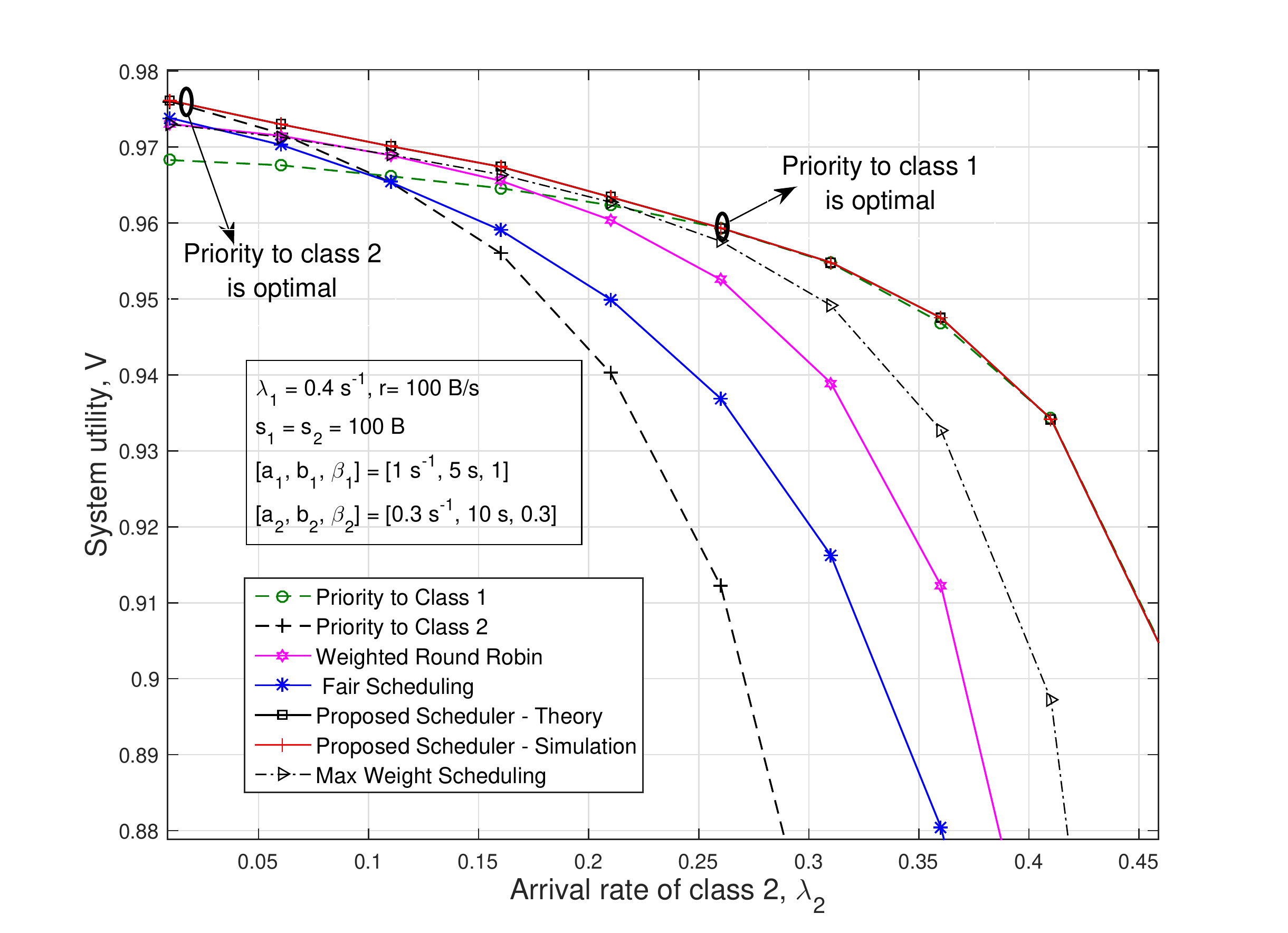}%
\caption{System utility for a mix of delay-sensitive and delay-tolerant traffic.}%
\label{sysUtlFig}%
\end{figure}

Fig.~\ref{sysUtlFig} shows the system utility performance of various packet schedulers as $\lambda_2$ is increased from $0.01~\text{s}^{-1}$ to $0.46~\text{s}^{-1}$. As expected, the system utility monotonically decreases with increase in $\lambda_2$ due to larger queuing delay at AS. We verify the correctness of the analytical result in Eq.~\eqref{finalSoln} as both the theoretical and simulation result match.

The proposed scheduler outperforms other schedulers with the performance gap being much larger at higher $\lambda_2$. This is because unlike other schedulers, it prioritizes service to delay-sensitive traffic at higher arrival rates of delay-tolerant traffic. Therefore, at very low $\lambda_2$, the proposed optimal scheduler converges to priority to class $2$, while at high $\lambda_2$ priority to class $1$ becomes optimal due to its delay-sensitive traffic. 

The queue size of both classes become comparable at high $\lambda_2$, and thus max-weight scheduling selects the packets from each class with roughly same frequency. Due to the increased server utilization by delay-tolerant traffic, the utility of delay-sensitive traffic and eventually the system utility is reduced. Similarly the performance of fair scheduler and WRR degrades at high $\lambda_2$ due to increased AS utilization by class $2$. The performance of WRR is bad despite assigning higher weight to delay-sensitive traffic due to the usage of non-optimal weights. Even with optimal weights, we still expect its performance to be sub-optimal as it aims to achieve (weighted) fairness of service between the delay-tolerant and delay-sensitive traffic. This is in direct conflict with service requirement of delay-sensitive traffic which needs to be prioritized over delay-tolerant traffic, particularly so at high arrival rate of delay-tolerant traffic. 
 
\subsection{Delay jitter performance of various schedulers}
\begin{figure}%
\centering
\includegraphics[height = 2.5 in, width = 3.3 in]{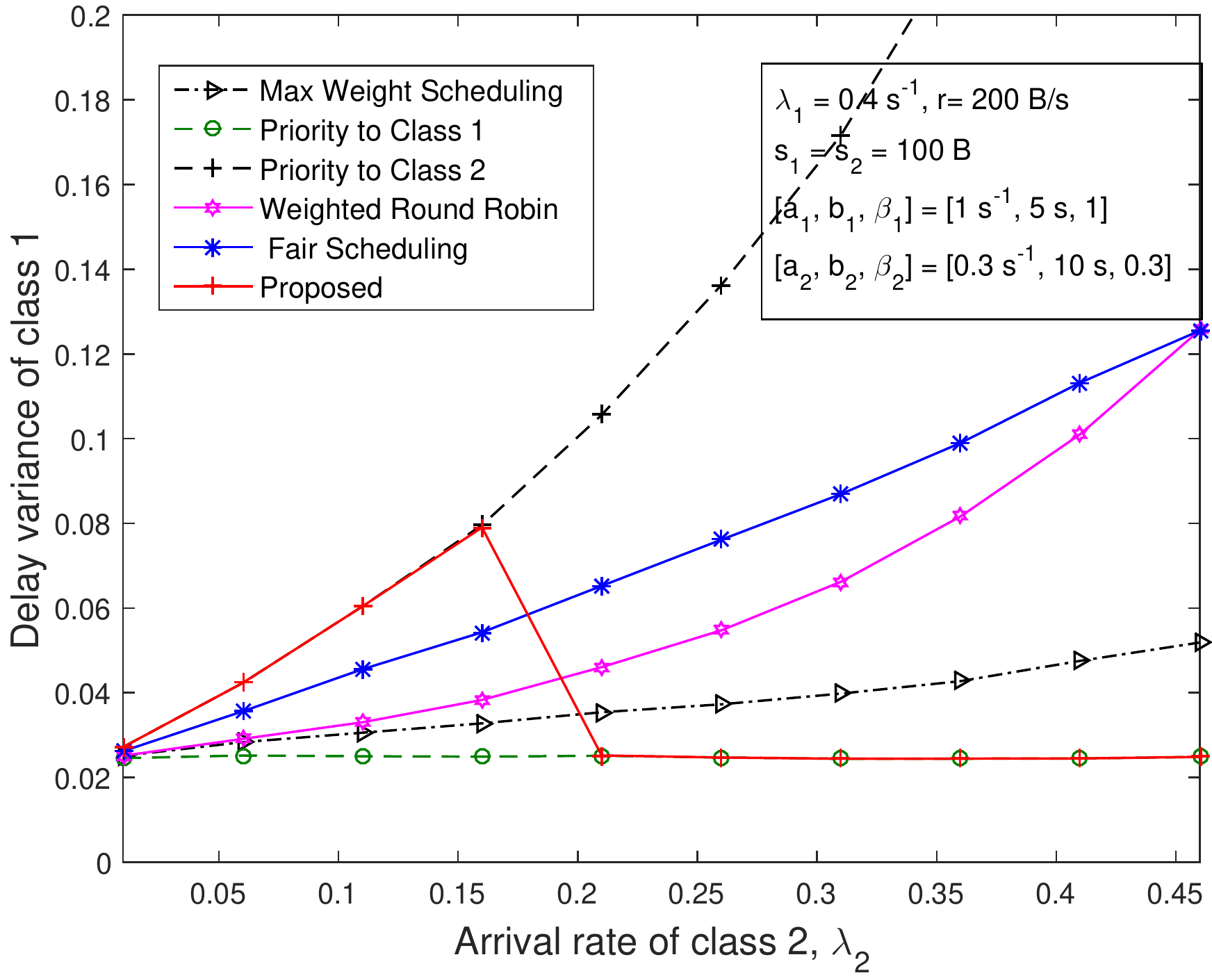}%
\caption{Delay variance of class $1$.}%
\label{delayVar1}%
\end{figure}

\begin{figure}%
\centering
\includegraphics[height = 2.5 in, width = 3.3 in]{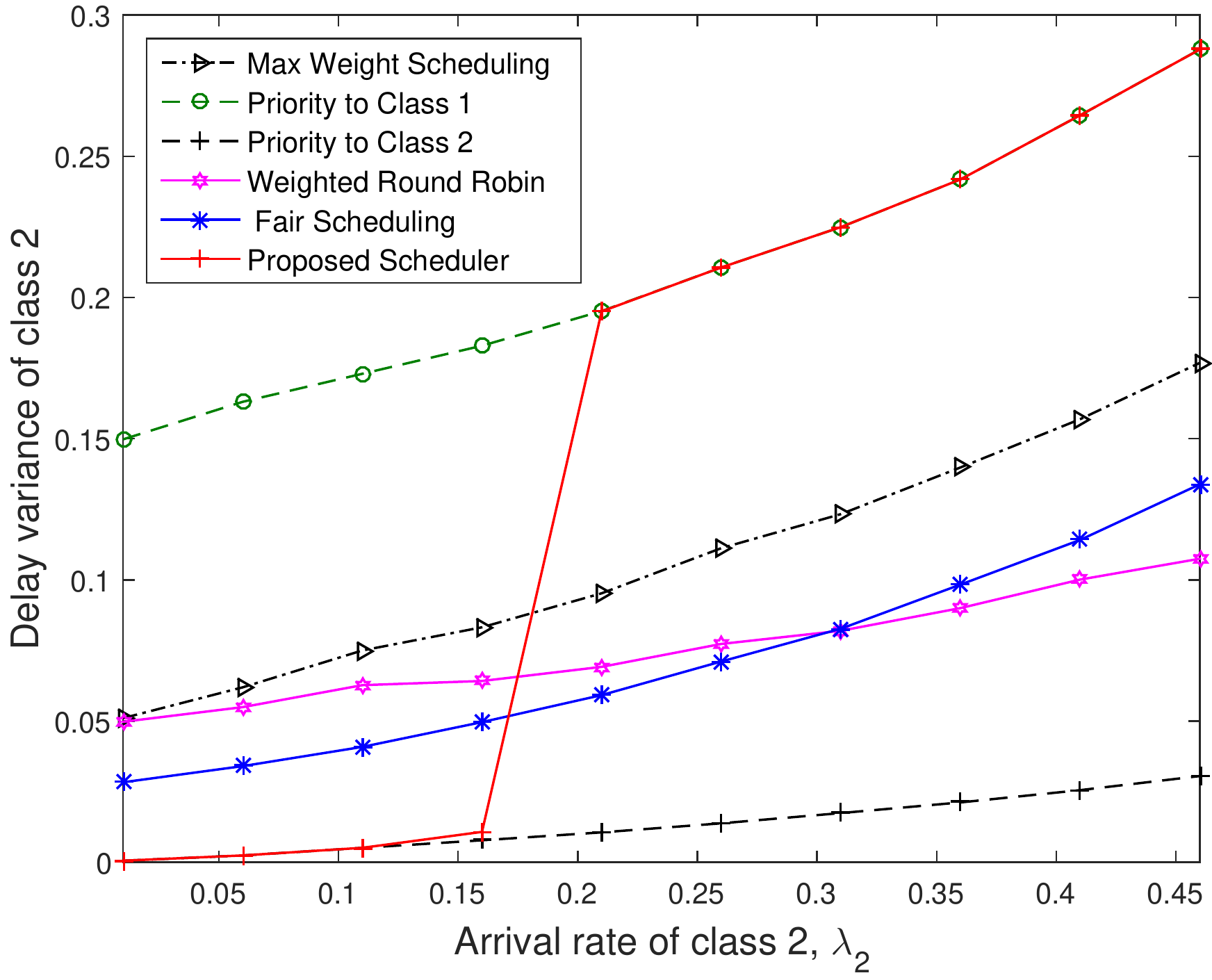}%
\caption{Delay variance of class $2$.}%
\label{delayVar2}%
\end{figure}

We now study the jitter performance by measuring the packet delay variance for each class under a given scheduling policy. Fig.~\ref{delayVar1} and \ref{delayVar2} show the plot of jitter of delay-sensitive class $1$ and delay-tolerant $2$ respectively as volume of delay-tolerant traffic ($\lambda_2$) is increased. We note that at high $\lambda_2$, the jitter for the delay-sensitive traffic is the least and same as that of priority-class $1$ scheduler. This is highly desirable for delay-sensitive applications. Furthermore, the jitter for delay-sensitive traffic is roughly constant as volume of delay-tolerant traffic is increased. Thus it is resilient to increase in delay-tolerant traffic. Although, the jitter of delay-tolerant traffic is quite high, it is usually acceptable due to its delay-tolerant nature and if needed, it can be mitigated using a jitter regulator. 

The transitions in the jitter for proposed scheduler in Fig.~\ref{delayVar1} and \ref{delayVar2} occur when it switches from one priority scheme to another (i.e., from $\alpha=0$ to $\alpha=1$). In fact by making the x-axis ($\lambda_2$) finer, we can allow the scheduler to take on all values of $\alpha$ between $0$ and $1$ as $\lambda_2$ is increased from $0.15$ and $0.2$.

\section{Conclusions and Future Work}
\label{concl}
In this paper, we presented a delay-optimal online packet scheduler for M2M uplink traffic. To this end, we first used a generic sigmoidal function to map the diverse latency requirements of delay-sensitive and delay-tolerant M2M traffic onto appropriate utility functions. We defined a system utility metric as the weighted product of average utility of two classes so as to ensure proportional fairness among them. The delay-optimal scheduler is obtained as a solution to the system utility maximization problem. We note that any work-conserving scheduling policy can be realized by appropriating time-sharing between the preemptive priority scheduling policies. Therefore, we obtain the optimal scheduling policy by solving for the optimal fraction of time-sharing between the two preemptive priority policies that results in maximum system utility. Using Monte-Carlo simulations for the queuing process at AS, we verify the correctness of the analytical result for optimal scheduler and compared its performance with various state-of-the-art schedulers. We show that the proposed scheduler outperforms other schedulers with the performance gap quite large at high arrival rate of delay-tolerant traffic. The proposed scheduler converges to either of priority scheduling policies as the arrival rate of delay tolerant traffic is varied. Lastly, we note that the delay jitter of delay-sensitive traffic at medium to high server utilization, is significantly lower than that of other scheduling policies, which is a highly desirable feature for delay-sensitive traffic. This comes at the expense of somewhat higher delay variance for delay-tolerant traffic which is usually acceptable due to its delay-tolerant nature.\\

\bibliographystyle{IEEEtran}

\bibliography{delayOptimal}

\begin{thebibliography}{10}
\providecommand{\url}[1]{#1}
\csname url@samestyle\endcsname
\providecommand{\newblock}{\relax}
\providecommand{\bibinfo}[2]{#2}
\providecommand{\BIBentrySTDinterwordspacing}{\spaceskip=0pt\relax}
\providecommand{\BIBentryALTinterwordstretchfactor}{4}
\providecommand{\BIBentryALTinterwordspacing}{\spaceskip=\fontdimen2\font plus
\BIBentryALTinterwordstretchfactor\fontdimen3\font minus
  \fontdimen4\font\relax}
\providecommand{\BIBforeignlanguage}[2]{{%
\expandafter\ifx\csname l@#1\endcsname\relax
\typeout{** WARNING: IEEEtran.bst: No hyphenation pattern has been}%
\typeout{** loaded for the language `#1'. Using the pattern for}%
\typeout{** the default language instead.}%
\else
\language=\csname l@#1\endcsname
\fi
#2}}
\providecommand{\BIBdecl}{\relax}
\BIBdecl

\bibitem{M2Mgrowth}
G.~Intelligence. (2014, Feb.) From concept to delivery: the m2m market today.
  https://goo.gl/yFmi5s.

\bibitem{smartGrid}
J.~J. Nielsen, G.~C. Madueño, N.~K. Pratas, R.~B. Sørensen, C.~Stefanovic,
  and P.~Popovski, ``{What can wireless cellular technologies do about the
  upcoming smart metering traffic?}'' \emph{IEEE Communications Magazine},
  vol.~53, no.~9, pp. 41--47, September 2015.

\bibitem{openSG}
E.~Hossain, Z.~Han, and H.~V. Poor, \emph{Smart Grid Communications and
  Networking}.\hskip 1em plus 0.5em minus 0.4em\relax Cambridge University
  Press, 2012.

\bibitem{Maia14}
A.~Maia, D.~Vieira, M.~de~Castro, and Y.~Ghamri-Doudane, ``{Comparative
  performance study of LTE uplink schedulers for M2M communication},'' in
  \emph{IFIP Wireless Days (WD)}, Nov 2014, pp. 1--4.

\bibitem{Gotsis12}
A.~Gotsis, A.~Lioumpas, and A.~Alexiou, ``Evolution of packet scheduling for
  machine-type communications over lte: Algorithmic design and performance
  analysis,'' in \emph{IEEE Globecom Workshop}, Dec 2012, pp. 1620--1625.

\bibitem{Buttazzo11}
G.~Buttazzo, \emph{{Hard Real-Time Computing Systems: Predictable Scheduling
  Algorithms and Applications}}.\hskip 1em plus 0.5em minus 0.4em\relax New
  York: Springer, 2011.

\bibitem{sysCon16arxiv}
\BIBentryALTinterwordspacing
A.~Kumar, A.~Abdelhadi, and C.~Clancy, ``{An Online Delay Efficient Packet
  Scheduler for {M2M} Traffic in Industrial Automation},'' \emph{CoRR}, vol.
  abs/1601.01348, 2016. [Online]. Available:
  \url{http://arxiv.org/abs/1601.01348}
\BIBentrySTDinterwordspacing

\bibitem{milcom16arxiv}
\BIBentryALTinterwordspacing
------, ``An online delay efficient multi-class packet scheduler for
  heterogeneous {M2M} uplink traffic,'' \emph{CoRR}, vol. abs/1601.03061, 2016.
  [Online]. Available: \url{http://arxiv.org/abs/1601.03061}
\BIBentrySTDinterwordspacing

\bibitem{Demers89}
A.~Demers, S.~Keshav, and S.~Shenker, ``Analysis and simulation of a fair
  queueing algorithm,'' in \emph{SIGCOMM}, 1989.

\bibitem{Greenberg90}
A.~Greenberg and N.~Madras, ``{How fair is fair queueing?}'' in \emph{Proc.
  Performance}, 1990.

\bibitem{Parkeh93}
A.~K. Parekh and R.~G. Gallager, ``A generalized processor sharing approach to
  flow control in integrated services networks: the single-node case,''
  \emph{IEEE/ACM Transactions on Networking}, vol.~1, no.~3, pp. 344--357, Jun
  1993.

\bibitem{Tassiulas}
L.~Tassiulas and A.~Ephremides, ``Stability properties of constrained queueing
  systems and scheduling policies for maximum throughput in multihop radio
  networks,'' in \emph{IEEE Conference on Decision and Control}, Dec 1990.

\bibitem{Dhillon14}
H.~S. Dhillon, H.~Huang, H.~Viswanathan, and R.~A. Valenzuela, ``{Fundamentals
  of Throughput Maximization With Random Arrivals for M2M Communications},''
  \emph{IEEE Transactions on Communications}, vol.~62, no.~11, pp. 4094--4109,
  Nov 2014.

\bibitem{AbdelhadiCNC2014}
A.~Abdelhadi and T.~Clancy, ``{A Utility Proportional Fairness Approach for
  Resource Allocation in 4G-LTE},'' in \emph{IEEE ICNC, CNC Workshop}, 2014.

\bibitem{AbdelhadiPIMRC2013}
------, ``{A Robust Optimal Rate Allocation Algorithm and Pricing Policy for
  Hybrid Traffic in 4G-LTE},'' in \emph{IEEE PIMRC}, 2013.

\bibitem{gallager}
D.~Bertsekas and R.~Gallager, \emph{Data Networks (2nd Ed.)}.\hskip 1em plus
  0.5em minus 0.4em\relax Prentice-Hall, Inc., 1992, ch. Delay Models in Data
  Networks, pp. 203--206.

\end{thebibliography}
\end{document}